\documentclass{ieeeaccess}
\usepackage{cite}
\usepackage{amsmath,amssymb}
\usepackage{graphicx}
\usepackage{booktabs}
\usepackage{array}
\usepackage{siunitx}
\usepackage{url}
\usepackage[hidelinks]{hyperref}
\input{tables/phaseb_values}

\begin{document}

\history{Received 10 October 2026, accepted 29 October 2026.}
\vol{14}
\year{2026}
\doi{10.1109/ACCESS.2026.3749922}

\title{Same Request, Different Answer: Quantization Amplifies Cache-Induced
Divergence in LLM Serving}

\author{\uppercase{Aditi Patodiya}}
\address{Independent Researcher, Milpitas, CA 95035 USA}

\markboth{A. Patodiya: Quantization Amplifies Cache-Induced Divergence in LLM Serving}
{A. Patodiya: Quantization Amplifies Cache-Induced Divergence in LLM Serving}

\corresp{Corresponding author: Aditi Patodiya (e-mail: aditi.patodia31@gmail.com).}

\begin{abstract}
Major open-source serving stacks reuse the key-value tensors of a shared
prefix by default. We measure what this prefix reuse costs in
reproducibility. Holding model, decoding, seed and request order
fixed at batch size one, we ran an eighty-episode multi-turn agentic tool-use
workload with prefix reuse on and off across two engines and four weight
formats. Without prefix reuse, repeated runs reproduced every
token and log-probability in all ten configurations. With it, cache-hit
and recompute paths diverged on both engines. On llama.cpp
with GGUF weights this path divergence rises with quantization: for
Qwen2.5-7B over ten independent server histories, it reached
\pbv{sweep.F16.mean} percent of episodes at 16-bit and \pbv{sweep.Q4.mean}
percent at four-bit. Near-tie decisions became only about \pbv{mech.near.ratio} times as common,
while the cache-path change in top-two margins grew
\pbv{mech.pert.ratio}-fold from 16-bit to three-bit. Serving four-bit weights through 16-bit kernels
left divergence not detectably different from 16-bit, placing the amplification
mainly in the quantized execution path. At Q4\_K\_M, run-to-run divergence came
mainly from llama.cpp's server-level prompt cache: under the canonical episode order it
moved divergence by 37.5 points, and at its default, divergence spanned
\pbv{bsix.min} to \pbv{bsix.max} percent under five random orders. With cache
state restored, each path reproduced exactly while the two still differed.
In the tested stacks, cached serving is irreproducible because the cache
state is absent from the request and not reset by default. A single-turn bridge shows
path divergence reaching task outcomes with no detectable net change in
accuracy. We
release the harness, logs, and analysis pipeline.
\end{abstract}

\begin{keywords}
Large language models, inference serving, reproducibility, prompt caching,
key-value cache, quantization, software testing, empirical software
engineering
\end{keywords}

\titlepgskip=-15pt

\maketitle

\section{Introduction}
\label{sec:intro}

Prefix caching is one of the least controversial optimizations in modern
language model serving. When consecutive requests share a leading span of
tokens, the server reuses the key and value tensors it already computed for
that span instead of recomputing them. Every major serving stack ships the
feature and most enable it by default \cite{kwon2023efficient,
zheng2024sglang, gim2024promptcache}. The savings are large and well
documented: a recent evaluation of long-horizon agent pipelines reports 41
to 80 percent lower cost and 13 to 31 percent faster time to first token
when caching is on \cite{lumer2026dontbreak}. For agentic workloads the
appeal is even stronger than for chat, because an agent re-sends its entire
growing transcript on every step, so almost the whole prompt is a cache hit
after the first turn.

The optimization carries an implicit promise. Reusing arithmetic that was
already performed is supposed to be an implementation detail, invisible
above the serving layer. A user who sends the same request with the same
sampling parameters expects the same answer whether or not the server
happened to have a warm cache.

That promise does not hold, and the failure is not subtle. Reusing cached
keys and values can change the order in which floating-point accumulations are
performed, and floating-point addition is not associative, so the reused
path and the recomputed path can produce slightly different numbers. Most of
the time the difference is invisible. Occasionally it moves a token
probability across a decision boundary, and the sampled token changes. In a
single-turn setting that is a curiosity. In an agent loop it is not, because
the changed token may be part of a tool call, and the tool call determines
the next observation, which determines the rest of the episode.

Practitioners have noticed. Two issues in the vLLM tracker report
behavior of this kind. One describes an accuracy drop from roughly 78
percent to 60 percent when the prefix cache interacts with the recomputation
path \cite{vllm2025issue18055}. The other, filed against ROCm hardware and
not reproducing on CUDA, shows an identical request returning different text
depending on whether it hit the cache \cite{vllm2026issue33123}. Both were
closed without a systematic characterization. In the literature, one recent study established that
cached and recomputed decoding are not numerically equivalent under FP16 on
single-turn arithmetic \cite{chodavarapu2026illusion}, and a study of
inference backends found that backend choice alone can shift benchmark
scores by as much as 16.6 points, naming prefix caching among the causes, although its ablation of that
feature measured aggregate accuracy alone \cite{pape2026silent}. What has been missing is a
controlled measurement of how often this matters for the workload the
feature was designed to accelerate, and of what the effect does when the
weights are quantized, which is how most locally served models run.

This paper provides that measurement. We run a paired design in which the
only variable is the cache setting, with greedy decoding, a fixed seed, a
batch size of one, and serial requests, so the usual sources of serving
nondeterminism are held out. The workload is multi-turn agentic tool use
from the Berkeley Function Calling Leaderboard \cite{patil2025bfcl}, with
single-turn grade-school mathematics \cite{cobbe2021training} as a bridge to
the prior single-turn result. We run every configuration twice, which turns
the design into something stronger than a comparison between arms: it lets
each arm be compared against itself.

That design produces the paper's central result. With prefix
reuse disabled, outputs from repeated executions of the identical workload
match exactly in every configuration we tested, down to the log-probability of
every token. With
prefix reuse enabled, two different things happen. The cache-hit path and the
recompute path produce different trajectories for a large share of episodes
(path divergence), and repeated cache-enabled runs disagree with each other when
the server carries state between them, mostly through a second, server-level
prompt cache that llama.cpp enables by default (run-to-run divergence). In the
configurations we tested, restoring the cache state that prefix reuse reads
removed all observed variation within each path, so the server
behaved deterministically given its state. What it lacks is reproducibility from the request alone, because the state that
decides the output is not part of the request.

The mechanism is not news, and we do not present it as such. Work on
deterministic inference already identifies reduction-order variation between
prefilled and cached execution as a source of nondeterminism, and prescribes
batch-invariant kernels to remove it \cite{he2025defeating}. SGLang now ships
a deterministic mode \cite{sglang2025deterministic} that reports consistent
output across cached and uncached prefill on two of its three attention
backends \cite{sglang2026determdocs}. What has not been established is the
magnitude in the configuration people actually deploy. Those fixes are
opt-in, they carry a throughput cost, they are unavailable on some backends, and neither llama.cpp nor vLLM
offered them in the versions we tested. Our contribution is to measure what the default
configuration does to the workload the cache was built for.

The contributions are as follows.

\begin{itemize}
\item A controlled measurement across model families, quantization formats,
and two serving implementations, showing that cache-disabled serving repeats
exactly, in every token and every reported log-probability, while
cache-enabled serving departs from it.
\item An isolation of the cause. Under the canonical episode order, a single
server-level prompt-cache setting is shown to move run-to-run divergence by 37.5
percentage points, and at its default the divergence ranges from
\pbv{bsix.min} to \pbv{bsix.max} percent under five random orders; once the
setting is disabled, the order in which passes run has no effect in the configuration tested. A reset control, repeated across weight
formats and independent server sessions, demonstrates that the cached path is
reproducible once cache state is restored. Under the configurations tested,
restoring cache state removed all observed variation within each path, and that
state is something deployments neither report nor control.
\item A replicated measurement of how weight quantization interacts with path
divergence for Qwen2.5-7B on llama.cpp with GGUF weights. Over ten independent server
histories, divergence rises from \pbv{sweep.F16.mean} percent of episodes at
F16 to \pbv{sweep.Q4.mean} percent at Q4\_K\_M, an odds ratio of
\pbv{sweep.or.hist} per format step (95\% CI \pbv{sweep.or.cilo} to
\pbv{sweep.or.cihi}). Direct top-two margin measurements, and a control that
serves the four-bit weights through the 16-bit kernels, point to the quantized
execution path rather than the rounded weight values as the main source of the
amplification.
\item An outcome-level analysis separating instability from degradation,
independently replicating a concurrent single-turn result in a
serving-engine setting: individual items change correctness in both
directions with no detectable net change in accuracy.
\item A released harness, raw per-request logs including server-reported
cache exposure, and an analysis pipeline that regenerates every number in
this paper from the released logs and telemetry summary.
\end{itemize}

\section{Background and Related Work}
\label{sec:related}

\subsection{Prefix caching in production serving stacks}

Key-value caching within a single generation is standard. What concerns us
here is reuse \emph{across} requests. PagedAttention introduced the memory
management that makes such sharing practical by storing the cache in
non-contiguous blocks that different sequences can reference
\cite{kwon2023efficient}. RadixAttention generalized the idea to a prefix
tree over live requests, which is a natural fit for agent loops where many
requests share long leading spans \cite{zheng2024sglang}. Prompt Cache went
further and reused attention states for non-contiguous prompt modules
\cite{gim2024promptcache}, and CacheBlend fused cached knowledge for
retrieval-augmented serving with selective recomputation
\cite{yao2025cacheblend}. Cache eviction and windowing schemes add another
layer of state that varies between runs \cite{zhang2023h2o,
xiao2024efficient}.

These systems are evaluated on throughput, latency, and memory, and often
on aggregate task accuracy. What they do not report is whether a given
request returns the same answer with a warm cache as with a cold one. The
closest evaluation to ours in workload terms measured caching across more
than five hundred agent sessions but recorded only cost and time to first
token \cite{lumer2026dontbreak}. The behavioral question was left open, and
the field guidance that resulted, namely to cache aggressively, was issued
without it.

\subsection{Nondeterminism in LLM inference}

Practitioners have long observed that temperature zero does not guarantee
identical output, and this has been documented systematically
\cite{atil2024nondeterminism}. The mechanism that has received the most
recent attention is batch invariance. Because reduction order in fused
kernels depends on how requests are grouped, a request batched with
different neighbors takes a different arithmetic path, and rewriting kernels
to be batch invariant removes that source \cite{he2025defeating}. SGLang
implemented such a deterministic mode as an opt-in flag
\cite{sglang2025deterministic}; it supports the radix cache on two of its
three attention backends \cite{sglang2026determdocs}, and its reported
performance benchmarks were run with that cache disabled, so the reported cost
of determinism excludes the cache. Neither source reports how often cached
and uncached prefill diverge in the default mode. A complementary line restores determinism
through verified speculation rather than kernel redesign
\cite{gond2026llm42}.

Our measurements deliberately exclude the cross-request part of the
batch-invariance story.
Every request in this study is issued alone, at batch size one, in serial
order, so grouping cannot vary between runs. The cache-disabled arms confirm
this directly by repeating exactly, token for token, under the same
conditions. What prefix reuse changes is how many of a request's tokens one
forward pass computes, the within-request case that batch-invariant kernels
must also cover \cite{he2025defeating}.

Two studies bear directly on the cached path. One demonstrated that FP16
decoding with a key-value cache is not numerically equivalent to
recomputation, using single-turn arithmetic and three models, and attributed
the effect to accumulation order under FP16 non-associativity
\cite{chodavarapu2026illusion}. Another quantified how much benchmark scores
move when only the inference backend changes, reporting shifts up to 16.6
points and naming prefix caching among the mechanisms
\cite{pape2026silent}. We extend the first study to
multi-turn agent episodes and add quantization as a factor. We complement the
second study's aggregate-accuracy ablation of prefix caching by measuring
divergence request by request and episode by episode.

\subsection{Quantization and behavioral change}

Quantization studies usually report aggregate accuracy, and by that measure
modern low-bit formats are close to lossless \cite{kurtic2025bf16}. A more
careful reading of compressed model behavior shows that this framing hides
churn: models that match a baseline on aggregate accuracy can disagree with
it on a substantial share of individual items \cite{dutta2024accuracy}.
Damage from quantization is also unevenly distributed across populations
that aggregate metrics average over \cite{marchisio2024quantization}.

That distinction is the one our results turn on. We do not claim that
quantization degrades accuracy, and our data would not support such a claim.
We report instead that quantization changes how strongly the cache path
perturbs the computation, and therefore how often that perturbation flips a
decision. That is a question about the numerical difference between two
arithmetic paths, not about the model's average competence.

\section{Study Design}
\label{sec:design}

\subsection{Research questions}
\label{sec:rqs}

\noindent\textbf{RQ1 (baseline determinism).} With prefix reuse disabled,
does repeated execution of an identical workload reproduce every emitted token
and its log-probability under greedy decoding on a fixed serving stack?

\noindent\textbf{RQ2 (cache-induced divergence).} With prefix reuse
enabled, how often does the output differ from the recompute path (path
divergence) and from a repeat of the same run (run-to-run divergence), and how
deep into an episode does the first difference appear?

\noindent\textbf{RQ3 (quantization interaction).} Does weight quantization
amplify or damp path divergence?

\noindent\textbf{RQ4 (backend generality).} Do the effects reproduce across
independent serving implementations?

\noindent\textbf{RQ5 (task outcomes).} Does cache-induced divergence change
task correctness, and if so, is the change directional or merely unstable?

\subsection{Two caching mechanisms and two effects}
\label{sec:terms}

Two mechanisms in this study keep state between requests, and they need
separate names. \emph{Prefix reuse} is the reuse of key and value tensors
already computed for the leading tokens of a prompt. llama.cpp controls it per
request through its \texttt{cache\_prompt} flag, and vLLM through its prefix
caching option. It is the variable we manipulate, and \emph{cache-enabled} and
\emph{cache-disabled} in this paper always refer to it. The \emph{server-level
prompt cache} is a second layer in llama.cpp: a host-memory store of whole
conversation states, enabled by default and controlled by the
\texttt{-{}-cache-ram} option (Section~\ref{sec:mechanism}). We name it
explicitly wherever it matters.

The effects we measure are also distinct. \emph{Path divergence} is a difference
between the cache-hit path and the recompute path for the same request.
\emph{Run-to-run divergence} is a difference between two runs of the same
workload with prefix reuse enabled, and it arises from state the server carried
over from earlier requests. Table~\ref{tab:terms} lists the four terms.

\begin{table}[!t]
\caption{Terms for the two caching mechanisms and the two effects.}
\label{tab:terms}
\centering
\small
\begin{tabular}{@{}>{\raggedright\arraybackslash}p{0.29\columnwidth}>{\raggedright\arraybackslash}p{0.63\columnwidth}@{}}
\toprule
Term & Meaning in this paper \\
\midrule
Prefix reuse & Reuse of cached key and value tensors for a shared prompt
prefix; the manipulated variable \\
Server-level prompt cache & llama.cpp's host-memory store of conversation
states; disabled in the replicated sweep and its other controls, at its default in
the production-default runs \\
Path divergence & Same request, cache-hit path against recompute path \\
Run-to-run divergence & Same workload rerun with prefix reuse enabled, differing
because of carried-over server state \\
\bottomrule
\end{tabular}
\end{table}

\subsection{Experimental protocol}
\label{sec:protocol}

The study uses a paired design. A \emph{cell} is one complete pass of a
workload through one serving configuration with prefix reuse either
enabled or disabled. Everything outside the caching setting is held fixed
within a configuration: the model weights, the quantization format, the
engine build, the sampling parameters, the request order, and the hardware.

Sampling is greedy throughout. Temperature is zero, the random seed is
fixed at 42, and every request carries the same seed. Requests are issued
serially with a batch size of one, so no request shares a forward pass with
another. This matters because continuous batching is itself a documented
source of nondeterminism, and leaving it active would confound the
measurement we are trying to isolate. The key-value cache is stored in
16-bit floating point in every arm, including the arms that serve quantized
weights, so that cache precision never varies with the weight format.

Two properties of the design carry the argument. First, in the original grid, each
configuration is run twice per arm, which yields a run-to-run comparison: the
same arm against itself. The cache-disabled run-to-run comparison is the internal
validity check, since any difference there would indicate a nondeterminism
source we failed to control. Second, the cache setting is verified from the
server rather than assumed wherever the engine permits it. On llama.cpp every
response records how many prompt tokens were served from cache, making cache
exposure a measured per-request variable rather than an inferred property of
the arm. vLLM 0.11.0 leaves that field unpopulated on the endpoint we use, so
there the check comes from the engine log instead, as Section~\ref{sec:rq4}
reports.

Cache control differs by engine and we follow each engine's own mechanism.
We use llama.cpp \cite{gerganov2023llamacpp} with GGUF weights
\cite{gguf2023spec} and vLLM. In llama.cpp prefix reuse is a per-request flag,
so in the original grid both arms run against a single server process, and in
the replicated sweep every pass runs on a fresh one (Section~\ref{sec:replication}). In vLLM prefix caching is set when the server is launched, so
each arm runs against its own server process launched with the corresponding
flag. A third engine, SGLang, entered the frozen study plan as a
budget-contingent stretch lane, which the plan specified would be the first
component dropped if the compute allowance required it. It was dropped under
that rule before any collection, so no SGLang cells exist and none are
reported. Engine versions, launch flags, and model checksums are recorded and
released with the artifact.

All measurements ran on rented NVIDIA RTX 4090 GPUs with 24\,GB of memory,
compute capability 8.9, under Ubuntu 24.04 with the CUDA 12.6 toolkit. The
serving stacks are llama.cpp built from source at release b10434, commit
7e4c0a9, with CUDA offload, and vLLM 0.11.0 on PyTorch 2.8.0+cu128 with
transformers 4.57. The grid of Table~\ref{tab:main} ran under NVIDIA driver
570.211.01, and the experiments of Section~\ref{sec:mechanism}, apart from the
controls of Section~\ref{sec:replication} reported there, ran under
580.126.20 on a second machine of the same type, so comparisons across those
two groups carry a driver difference. Every comparison from which we
draw a causal conclusion was collected within a single machine and driver
version, with one exception that we checked directly. One cell of the pass-order
comparison comes from the original grid. Its first cache-enabled pass is
identical, in every token and log-probability, to the first pass of the new run
it is compared with. The replicated sweep and
the controls of Section~\ref{sec:replication} ran on a third machine of the same
type under driver 595.91.07 with the same llama.cpp build. Before collecting
them we checked that this machine reproduces the earlier measurements: a
cache-disabled and a cache-enabled pass of Qwen2.5-7B at Q4\_K\_M matched the
corresponding earlier passes in every emitted token and every reported
log-probability, in 80 of 80 episodes each. The cache-disabled reference passes
of the sweep, recorded with five log-probabilities, also reproduced the earlier
cache-disabled grid token for token in all four formats, and the reference of
each format was identical in every token and log-probability under two
different episode orders.
Table~\ref{tab:settings} lists the principal
settings that can affect the measurements and the reason for each value.

\begin{table*}[!t]
\caption{Settings that can affect the measurements. Nothing is trained or
tuned on outcomes; the table gives the reason for each value.}
\label{tab:settings}
\centering
\small
\begin{tabular}{@{}>{\raggedright\arraybackslash}p{0.20\textwidth}>{\raggedright\arraybackslash}p{0.36\textwidth}>{\raggedright\arraybackslash}p{0.38\textwidth}@{}}
\toprule
Setting & Value & Reason \\
\midrule
llama.cpp build & release b10434, commit 7e4c0a9, CUDA 12.6, sm\_89 & cache behavior is version-specific \\
vLLM build & 0.11.0, PyTorch 2.8.0+cu128, transformers 4.57 & same \\
Slots and batching & one slot, batch size 1, serial requests & removes batch-composition effects \\
Context length & 16,384 tokens & holds every episode but one: in one random order of the production-default reruns, one episode exceeded it at the same request in both passes (Section~\ref{sec:mechanism}) \\
Key-value cache precision & 16-bit floating point in every arm & keeps cache precision independent of the weight format \\
Server-level prompt cache & default (8192\,MiB) in the original grid, the GSM8K bridge and the production-default reruns; both settings in the flag, order and latency experiments; disabled in the replicated sweep and its other controls & the variable isolated in Section~\ref{sec:mechanism} \\
Decoding & greedy, temperature 0, seed 42 on every request & removes sampling as a source of variation \\
Maximum new tokens & 4096 (BFCL, capped by remaining context); 1024 (GSM8K) & reached by \pbv{cap.bfcl.n} of \pbv{cap.bfcl.total} agentic and \pbv{cap.gsm8k.n} of \pbv{cap.gsm8k.total} single-turn requests across all runs \\
Reported log-probabilities & top 1 (original runs, production-default reruns, latency runs); top 5 (replicated sweep and its isolation, reset and dequantization controls) & top 5 gives the exact top-two margin; emitted tokens are unchanged by it \\
Workload selection & first 80 BFCL multi-turn base episodes by ID; GSM8K test items 0--499 & fixed before inspecting results \\
Bootstrap, single pass & 10,000 replicates; circular blocks; length by the Politis--White rule, at least 5 (realized 5 to 13); seed 20261007 & dependence-aware intervals \\
Bootstrap, replicated sweep & episodes and histories resampled independently; 10,000 replicates (\pbv{sweep.or.reps} with model refits for the odds ratio) & carries episode and history variation \\
Mixed model & random intercept per episode; formats coded 0 to 3 in the order F16, Q8\_0, Q4\_K\_M, Q3\_K\_M; adaptive Gauss--Hermite quadrature, 100 nodes & paired trend across formats \\
History orders & 10 random permutations, order $k$ drawn with seed (20261007, $k$), shared by all formats & pairs formats on episode and history \\
\bottomrule
\end{tabular}
\end{table*}

\subsection{Workloads}
\label{sec:workloads}

The primary workload is multi-turn agentic tool use. We use the multi-turn
base category of the Berkeley Function Calling Leaderboard, which places a
model in a simulated environment of stateful APIs and scores whether the
resulting environment state and call sequence match a reference. Episodes
run for several user turns, and within each turn the agent may take many
steps, so a single episode issues about ten requests on average, ranging from
seven to thirteen across configurations, and
accumulates a context in the thousands of tokens. That growth pattern is
what makes the workload appropriate here: each step re-sends the whole
conversation, which is exactly the shape of request that prefix caching is
designed to accelerate. Episodes are selected by a rule fixed before any
result was inspected, namely the first eighty entries in ascending
numeric identifier order.

The secondary workload is single-turn grade-school mathematics from GSM8K.
It serves two purposes. It provides a bridge to prior work on cached
inference, which studied single-turn arithmetic, and it supplies an outcome
measure clear of the floor: the models we run solve roughly nine in ten of
these problems, which still leaves enough incorrect items for flips to be
observable in both directions. Each item is presented four times in a
fixed order: twice on the recompute path and twice on the cache-hit path.
The repeated passes on each path give a per-path determinism check, so a
difference between paths is only counted as cache-attributable when each
path agreed with itself.

\subsection{Isolating the cause}
\label{sec:isolation}

Our first grid measured divergence but did not establish what produced it,
and a measurement of that kind invites two objections: that the passes
differed in execution order, and that comparing two runs whose cache states
differ is definitionally guaranteed to show a difference. Three further
experiments address both, all on Qwen2.5-7B at Q4\_K\_M under llama.cpp,
the configuration with the highest run-to-run divergence in the grid; the third
is also repeated for all four formats alongside the replicated sweep
(Section~\ref{sec:replication}).

The first varies execution order alone. One arm places a complete
cache-disabled pass between the two cache-enabled passes, matching the order
used in our llama.cpp lane; the other runs the two cache-enabled passes
adjacently, matching the vLLM lane. Everything else is held constant,
including a fresh server for each arm.

The second varies one flag. llama.cpp maintains a server-level prompt cache
(Section~\ref{sec:terms}) that is separate from prefix reuse, stores whole conversation
states, and selects among them by prefix similarity rather than by identity. We run the same procedure with that
layer disabled and at its default, again with a fresh server for each arm,
so the flag is the only difference.

The third re-establishes the cache state directly. Each item is served four times,
twice on the recompute path and twice on the cache-hit path, with the cold
state re-established before each recompute pass. Comparing each path against
itself under a restored state separates dependence on cache state from
residual randomness in either path. We do not use the engine's cache-erase
endpoint for this: it returns success while leaving the cached state intact,
which we verified from the reported cached-token counts, so the cold state is
established with the per-request flag that forces recomputation, whose
effect is visible in the same telemetry.

\subsection{Replicated controlled sweep}
\label{sec:replication}

The quantization comparison needs replication at the level where state lives. A
fresh server process given the same episode order reproduced its output token
for token in every check we ran: a second, independently started process at
the canonical order matched the first in all four formats, in every one of
\pbv{bone.tokens} tokens and their log-probabilities. What differs
between deployments is the history a server has processed, and with the
server-level prompt cache disabled, that history reaches an episode only through
the prefix that the preceding episode left in the cache. We vary it by
randomizing the episode order.

For each of the four Qwen2.5-7B formats, F16, Q8\_0, Q4\_K\_M and Q3\_K\_M, the
server-level prompt cache is disabled from launch and every pass runs on a fresh
server. Each format runs under eleven histories: the canonical ascending order
and ten random permutations drawn with a recorded seed. The same orders are used
for every format, so formats are paired on both episode and history, and a
Williams design balances the order in which formats run on the machine. Each
cache-enabled pass is compared with a cache-disabled reference pass of the same
format. The reference should not depend on history, and we check that rather
than assume it by running it under two different orders. Every request in these
runs records the five most probable tokens with their log-probabilities, which
gives the exact logit margin between the top two candidates at each position.

Five controls accompany the sweep. A second server process at the canonical
history checks process-level reproducibility in every format. An
episode-isolated variant overwrites the cache with a fixed sentinel prompt
before each episode, so that every episode starts from the same verified state,
and runs the workload under three orders. The restored-state control of
Section~\ref{sec:isolation} is repeated for all four formats in three
independent server sessions of 100 items each. The production-default
configuration, with the server-level prompt cache enabled, is rerun under five
random orders with both passes on one server, which measures how much the
run-to-run divergence it produces depends on history. Finally, to separate the
quantized weight values from the arithmetic that serves them, the Q4\_K\_M
weights are also dequantized to a 16-bit file, which keeps their values to
within 16-bit rounding but runs them through the F16 kernels, and the sweep is
repeated for that file under four histories.

\subsection{Measures}
\label{sec:measures}

Divergence is measured on token identifiers rather than on decoded text
wherever the engine returns them, so that a difference is registered even when
detokenization hides it. llama.cpp returns token identifiers on every
response. vLLM 0.11.0 returns token strings on the endpoint we use, so its two
configurations are compared on decoded tokens, which is a slightly weaker
criterion; each configuration records which criterion applied. For a
pair of runs we walk the request sequence of each episode in lockstep and
record the first position where the two runs differ, either in the prompt
that was assembled or in the emitted token sequence. An episode is counted
as divergent if any request differs or if the two runs issue different
numbers of requests, the latter meaning the agent took a different number
of steps.

Agentic outcomes are scored by the benchmark's own checker without
modification, so scoring is independent of our instrumentation; the checker
compares the final environment state and the sequence of invoked methods
against the reference. For the mathematics workload an answer is correct when
the final numeric value, extracted as described in Section~\ref{sec:rq5},
matches the reference.

Episodes within a pass run in sequence against one server, and this study's own
claim is that state carries across requests, so episodes are not
exchangeable and Wilson intervals may be anticonservative. Divergence rates
measured in a single pass, those of Tables~\ref{tab:twobytwo}
and~\ref{tab:main}, of Fig.~\ref{fig:mechanism} and of the production-default
reruns of Section~\ref{sec:mechanism}, carry a dependence-aware interval: a circular block bootstrap of the
episode-level indicators in execution order
\cite{politis1992circular}, with 10,000 replicates and the block length chosen
by the automatic rule of Politis and White \cite{politis2004automatic,
patton2009correction}, floored at five episodes. Blocks wrap around the end of
the sequence. We report the envelope of this interval and the Wilson interval,
so that allowing for dependence can widen an interval but never narrow it.
Where serial dependence is weak the block bootstrap comes out narrower, and the
reported interval is then the Wilson one. Realized block lengths run from 5 to
13 episodes, with most series at the floor, and recomputing every interval at
block lengths of 1, 5, 10 and 20 changes no conclusion. These single-pass
intervals are approximate: with 80 ordered episodes their coverage depends on the strength and form of the dependence, which is one reason the
replicated sweep rather than any single pass carries the quantization result. Rates in the replicated
sweep (Table~\ref{tab:sweep}, Fig.~\ref{fig:gradient}) and the differences
reported for its dequantization control use a
bootstrap that resamples episodes and histories independently, the two-way
design for crossed factors \cite{owen2007pigeonhole}, because each of those
quantities pools passes over several histories. Single-history values given
beside them, the canonical column of Table~\ref{tab:sweep} and the sentinel
rates, are descriptive. Results that describe exact
reproduction are reported as counts. Token counts in those results are counts of the log-probability records the
engine returns; llama.cpp merges a token that ends
inside a multi-byte character with the next one, so a record can cover more than
one token, and identical records imply identical response text. Perturbation comparisons use passes
recorded at the same top-$k$ depth, because llama.cpp's reported
log-probabilities depend in their last bits on that depth: requesting the top
five changes the order in which the reporting softmax is summed, not the logits
or any emitted token.
When no episode diverges, the bootstrap is degenerate and we report the count
itself; where a figure draws a bar for such a count, its caption names the
reference interval shown. We also test directly for position dependence, and report it in Section~\ref{sec:mechanism}.

The quantization trend is estimated with methods that keep each episode paired
across formats, with the formats entered as an ordinal score from 0 (F16) to 3
(Q3\_K\_M): a logistic model with a random intercept per episode, fitted by
maximum likelihood with adaptive quadrature; on the original grid, a
conditional logistic regression stratified by episode, fitted by conditional
maximum likelihood; and, as a secondary check, an exact test that permutes divergence across
formats within each episode, which assumes that the format labels are
exchangeable within an episode \cite{agresti2013categorical}. In the replicated sweep an
episode recurs in every history, so it remains the unit of the permutation
test: the test acts on each episode's divergence counts summed over histories
and permutes formats within the episode, never treating an episode's histories
as independent. In both cases the statistic is the sum of the format ranks over
all observed divergences, and its null distribution, with the formats
exchangeable within each episode, is computed exactly by convolution over
episodes. The random-intercept model is fitted to each history, its
slope is averaged over the histories, and the interval for that average comes
from the two-way bootstrap with the model refitted in every resampled history.
The odds ratio per format step summarizes the trend across the four formats.
The model treats the formats as ordered steps of equal size.
It does not claim that each step removes the same amount of precision.
Effect sizes, as paired differences in divergence rate, are reported first and
the tests second. In the logit-margin analysis of Section~\ref{sec:rq3},
per-episode summaries are compared across the ordered formats with Page's trend
test, with episodes as blocks, and ratios of format means carry intervals from a
bootstrap over episodes.

Paired outcome changes on GSM8K are tested with an exact McNemar test
\cite{mcnemar1947note} on the discordant pairs, which is the appropriate test
when the same item is observed under both conditions and we care whether
changes favor one condition \cite{dietterich1998approximate}.
The same problems appear in several configurations, so the pooled analysis is
done at the problem level, with a problem-clustered sign test and a bootstrap
that resamples problems. We report the discordant counts alongside the
p-value, since with small numbers of flips the counts themselves are more
informative than the test statistic.

\section{Results}
\label{sec:results}

\subsection{RQ1: cache-disabled execution repeats exactly}
\label{sec:rq1}

We begin with the control, because everything else depends on it. For every
configuration we executed the full eighty-episode workload twice with prefix
reuse disabled and compared the two runs request by request, on token
identifiers for llama.cpp and on token strings for vLLM.

Not one episode differed. In every configuration, across both serving
engines, across weight formats from 16-bit floating point down to three-bit
k-quantization, the second run reproduced the first exactly: identical
prompts at every step, identical emitted tokens, identical numbers of
requests per episode. The identity holds below the token level as well:
every emitted token carried an identical log-probability in both runs,
277,280 tokens across the ten configurations. On vLLM, which reports token strings
rather than identifiers, this covers the strings and their log-probabilities.
Table~\ref{tab:main} reports \num{0.0} in the cache-disabled column for every
configuration. We report this as an observed count (no divergent episode in any of the ten
configurations, 0 of 800 configuration-episode executions) rather than as a
pooled bound, because the configurations share episodes and are not
independent trials.

The value of this result is what it licenses. Under our conditions, greedy decoding with a fixed
seed, a batch size of one, and serial requests, the serving stack behaved as a
deterministic function of its input in every repetition we ran. Any
divergence observed elsewhere in this paper therefore cannot be attributed to
sampling, to scheduler nondeterminism, to batch composition, or to
unspecified engine noise, because all of those are equally present in the
cache-disabled arms and produced no observed variation. We also checked
every episode log for failures: no episode in any cell of the original grid raised an error, so
divergence cannot be an artifact of timeouts or exceptions.

\subsection{RQ2: cache-enabled execution diverges, and why}
\label{sec:rq2}

We then repeated the identical procedure with prefix reuse enabled,
changing nothing else about the workload or the decoding parameters.

Two distinct measurements follow, and separating them is the substance of
this section. Path divergence compares the cache-enabled arm against
the cache-disabled arm. Run-to-run divergence compares the cache-enabled arm
against itself, rerun.

\subsubsection{Cache-enabled and cache-disabled execution differ}

The path-divergence column of Table~\ref{tab:main} reports the first
comparison. In the production-default grid, enabling prefix reuse
changes the agent's trajectory on between 36.2 and 91.2 percent of episodes,
depending on configuration. This
measurement is stable: on the canonical history of the replicated sweep
(Section~\ref{sec:rq3}), run with the server-level prompt cache disabled, path
divergence for the four Qwen2.5-7B formats is \pbv{sweep.F16.canon},
\pbv{sweep.Q8.canon}, \pbv{sweep.Q4.canon} and \pbv{sweep.Q3.canon} percent,
against 36.2, 61.3, 75.0 and 77.5 percent in the grid.

The presumed mechanism is the familiar one. A request that hits the cache
reads keys and values for the shared prefix instead of recomputing them, so the
rest of the computation runs with different batch shapes and can accumulate in
a different order. Floating-point addition is not associative, and the resulting logits can differ in their low-order bits. We did not instrument the
kernels, so this account is an interpretation. What the measurement adds is
that the difference is not confined to low-order bits of the output. It
crosses token decision boundaries often enough to change agent behavior in a
large fraction of episodes. Divergence also begins early: across
configurations the first differing request has a median index between 1 and 4
of roughly ten, and within that request the first differing token has a median
index between 4 and 17. Among episodes that diverge, the first difference typically appears in the
opening exchanges, leaving several subsequent requests through which it can
propagate.

\subsubsection{On llama.cpp, a second caching layer drives run-to-run
divergence}

Rerunning the cache-enabled arm against itself (run-to-run divergence)
changed the trajectory on 8.8 to 77.5 percent of episodes across the ten
configurations of Table~\ref{tab:main}. Taken alone, that invites the
conclusion that cached serving is intrinsically nondeterministic.

It is not, and the next section shows why. The eight llama.cpp configurations
were collected with the server-level prompt cache left at its default, a
second caching layer, a host-memory prompt cache that is distinct from prefix
reuse and that persists whole conversation states across requests
independently of it.
The two vLLM configurations have no such layer; their run-to-run divergence
comes from prefix-cache state held in one server process
(Section~\ref{sec:mechanism}). Once the llama.cpp layer is disabled, repeated
cache-enabled runs of Qwen2.5-7B at Q4\_K\_M agree on 79 of 80 episodes. The
exception is the first episode run, which the first pass serves from an empty
cache and the second from the prefix the first pass left behind. In that
configuration the cached path is reproducible; what is not reproducible is
the state it reads from.

\subsection{The mechanism: state carried between runs}
\label{sec:mechanism}

\begin{figure}[!t]
\centering
\includegraphics[width=\columnwidth]{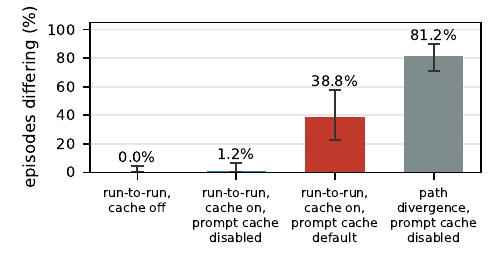}
\caption{Where the divergence comes from. Repeating a cache-disabled run
changes nothing. Repeating a cache-enabled run (run-to-run divergence) changes
little once the server-level prompt cache is disabled, and a great deal at its
default setting. Path divergence compares the cache-enabled and cache-disabled arms of the same run and has the highest observed rate shown. All four bars come from the controlled runs of
Section~\ref{sec:mechanism} rather than from Table~\ref{tab:main}. Bars show
dependence-aware 95\% intervals (Section~\ref{sec:measures}); for the zero
count the bootstrap is degenerate and the Wilson interval is shown. Qwen2.5-7B
at Q4\_K\_M under llama.cpp, 80 episodes per bar.}
\label{fig:mechanism}
\end{figure}

Three experiments isolate the cause. The first two use Qwen2.5-7B at
Q4\_K\_M, the configuration with the highest run-to-run divergence in the
grid, and hold
the model, episode set, decoding parameters, and server freshness constant.
The third re-establishes the cache state directly and is repeated in all four weight
formats. The vLLM
comparison at the end of the section uses the same model at BF16, the
checkpoint's native 16-bit format, since vLLM serves the unquantized weights.

Fig.~\ref{fig:mechanism} summarizes the result.

\textbf{A second caching layer controls run-to-run divergence.} Beyond the per-request
prefix reuse under study, llama.cpp maintains a server-level host-memory prompt cache that
stores whole conversation states, selects among them by longest-common-prefix
similarity rather than by identity, and evicts oldest-first
\cite{llamacpp2025hostcache}. It is enabled by default at 8192\,MiB and
controlled by a documented flag \cite{llamacpp2026serverreadme}, and it was
introduced in October 2025, so evaluations on llama.cpp builds from that date
with default settings could have been affected without notice. Our first grid
left it at its default. We held everything else fixed, including a fresh server for each arm and the
canonical episode order, and changed only that flag. Repeated cache-enabled
runs then diverged on 1 of
80 episodes with the layer disabled (1.2 percent, dependence-aware 95\% CI 0.0
to 6.7) and 31 of 80 with it at its default (38.8 percent, 22.5 to 57.5). Under
that order, changing this one setting produced a paired difference of 37.5
percentage points in run-to-run divergence (dependence-aware 95\% CI 21.2 to
57.5). We state the difference rather than the
ratio, since a single divergent episode in the disabled arm cannot pin a ratio
down.

The setting and its default are documented, and the documentation separately
warns that the per-request prefix cache can produce nondeterministic results
\cite{llamacpp2026serverreadme}. What we did not find in the documentation, in
the pull request that introduced the feature, or in the issue tracker is any
report that this second layer is an independent source of run-to-run variation
on an identical workload. Our contribution here is the measurement rather than
the discovery of the setting: a request's arithmetic path depends on which
stored state the similarity search selects, and that depends on what the server
processed earlier.

\textbf{Pass order acts through the same layer.} Our llama.cpp and vLLM
lanes had run their passes in different orders, so we tested order against the
flag above. Table~\ref{tab:twobytwo} assembles the four cells. Three are new
runs made for this purpose; the fourth, at the default setting with a
cache-disabled pass in between, is the corresponding configuration from
Table~\ref{tab:main}, whose first cache-enabled pass we verified is identical
to that of the new run in the same row. We label it as such rather than
presenting the table as a single designed experiment.

With the server-level prompt cache active, interposing a complete cache-disabled pass
between the two cache-enabled passes raises divergence from 38.8 to 77.5
percent, a difference of 38.8 points (dependence-aware interval 22.5
to 55.0), consistent with the intervening traffic rewriting the state that the
second pass inherits. With the layer disabled the same manipulation changes
nothing, 1.2 percent either way, and the two orderings produced identical
output, token for token, in every cell. Order therefore appears to act through the state the
layer carries rather than independently of it. The two rows do not share a
common starting point: at the default setting the first pass issues 854
requests and with the layer disabled it issues 827, so the setting changes
behavior within a pass as well as between runs.

Divergence in that configuration is also concentrated early in the run. Split
into quartiles of execution order, the counts of divergent episodes are 16, 6,
6 and 3, a trend that a permutation test puts at $p = 0.0002$. The same split in the
grid configuration shows no trend ($p = 0.59$). A layer that accumulates and
evicts entries as a run proceeds would be expected to produce exactly this
settling behavior, and it is a further reason to treat episodes within a cell
as dependent rather than as independent trials.

Those runs used one fixed episode order, so we repeated the production-default
configuration under five random orders, the first five of the replicated sweep,
with both cache-enabled passes back to back on one server, as in the
adjacent-passes column of Table~\ref{tab:twobytwo}. Run-to-run divergence averaged \pbv{bsix.mean} percent, with a
between-order standard deviation of \pbv{bsix.sd} points and a range from
\pbv{bsix.min} percent (dependence-aware 95\% CI \pbv{bsix.min.lo} to
\pbv{bsix.min.hi}) to \pbv{bsix.max} percent (\pbv{bsix.max.lo} to
\pbv{bsix.max.hi}). The canonical order, measured on
the second machine, gave 38.8 percent, above all five random orders measured on
the third machine. Although the third machine reproduced earlier Q4\_K\_M cache-disabled and
cache-enabled passes exactly, with the server-level prompt cache disabled
(Section~\ref{sec:protocol}), this comparison, made with that cache at its
default, does not isolate episode order from machine and driver differences.
The five random orders on the third machine, under otherwise fixed conditions,
show that the effect is present under every tested order and that its size
depends on the history the server has processed. The early concentration
holds as well: in all five orders the first quarter of the run contains the
most divergent episodes, with a permutation $p$ of at most
\pbv{bsix.trendp.max} in each. One of these orders was run twice on fresh
servers, and the two runs agreed in every one of \pbv{bsix.rep.tokens} tokens
and log-probabilities across both passes. In that order, then, with the server-level prompt cache at its default, the
server was deterministic given its history, and the
run-to-run divergence came from the state it carried from the first pass into
the second.
One episode in that order grew beyond the 16,384-token context window at the
same request in both passes, and it was compared on the requests both passes
completed.

\textbf{With cache state restored, the cached path is reproducible.} The reset
control serves each item four times, twice on the recompute path and twice on
the cache-hit path, re-establishing the cold state before each recompute pass
and verifying it from the reported cached-token counts on every request. In
the original control, 40 items in one session at Q4\_K\_M, the recompute path
reproduced on 40 of 40 items and the cached path on 40 of 40, while the two
paths differed from each other on 14 of 40. We repeated the control in all four
weight formats, in three independent server sessions each, with 100 items served in a different order in each session (Table~\ref{tab:reset}).
The telemetry confirmed the intended state on every request: all
\pbv{reset.ncold} recompute requests reported \pbv{reset.coldmax} cached tokens,
and every one of the \pbv{reset.nwarm} cache-hit requests reported at least
\pbv{reset.warmmin}. In all twelve sessions both paths reproduced on every
item. The two paths differed on \pbv{reset.F16.s1.differ} of 100 items at F16,
\pbv{reset.Q8.s1.differ} at Q8\_0, \pbv{reset.Q4.s1.differ} at Q4\_K\_M and
\pbv{reset.Q3.s1.differ} at Q3\_K\_M, with the same count in all three sessions
of each format. Within each path, every item's emitted tokens were identical
across the three sessions, although each session served the items in a
different order.

The same holds for whole episodes. When the cache is overwritten with a fixed
sentinel prompt before every episode, verified by zero cached tokens on each
episode's first request, every episode starts from the same state whatever ran
before it. Under three different episode orders, the canonical order and the
first two random orders of the sweep, the cache-enabled pass was then identical across the three orders in all four formats, in every one of \pbv{sent.tokens} compared tokens
and their log-probabilities. The cache-hit path still departed from the
recompute path on \pbv{sent.F16.crosspct}, \pbv{sent.Q8.crosspct},
\pbv{sent.Q4.crosspct} and \pbv{sent.Q3.crosspct} percent of episodes at F16,
Q8\_0, Q4\_K\_M and Q3\_K\_M. Isolation changes the state each episode
starts from, so these rates need not match the sweep's: Q8\_0 and Q3\_K\_M fall
outside the range of its histories, while every quantized format remains above
F16. With the history removed, path divergence and
its amplification by quantization remain. These
controls ran with the server-level prompt cache disabled, so they establish that the
cached path is a function of cache state in the regime where state is
otherwise controlled; they do not by themselves speak to the default
configuration, which the first two experiments cover.

The second engine points the same way through a different lever. vLLM exposes
no equivalent of the llama.cpp setting, so what varies there is server
lifetime. The configuration that launched a fresh server for each
cache-enabled pass diverged on 0 of 80 episodes, since each pass then begins
from an empty cache. Configurations that served both passes from one process diverged on 7 of 80
and 22 of 80 for Qwen2.5-7B in two sessions, and on 8 of 80 for Llama-3.1-8B
in its one session. The direction is consistent across every session; the
magnitude for the repeated model is not, 8.8 against 27.5 percent. Those two
sessions differ in the script that launched them, in whether an explicit GPU
memory fraction was set, and in the machine and driver version, and we did not
isolate which of these matters. We therefore take the vLLM lane as
establishing the direction of the effect on a second engine and the llama.cpp
lane as fixing its size.

Taken together, these results support a more precise claim than the raw
run-to-run numbers suggest. In the configurations tested, restoring cache
state eliminated all observed variation within each path, so cached serving
behaved as a deterministic function of the request \emph{and} the cache state.
Cached serving is irreproducible in practice because
cache state is not part of the request, is reported in the response at most
as a count of reused tokens, and is not reset between runs by default in any stack we tested. Disabling prefix
reuse removes the state, and with it both kinds of divergence. That gap, between a system that
is deterministic given its state and one that is reproducible from the
request, is what these experiments measure.

\begin{table}[!t]
\caption{Pass order and the server-level prompt cache, crossed. Values are
the fraction of 80 episodes whose trajectory changed when the cache-enabled arm
was repeated, with dependence-aware 95\% intervals. Order matters only while
that cache is active. Qwen2.5-7B at Q4\_K\_M under llama.cpp.}
\label{tab:twobytwo}
\centering
\small
\begin{tabular}{@{}lcc@{}}
\toprule
Server-level & passes & cache-off pass \\
prompt cache & adjacent (\%) & in between (\%) \\
\midrule
default & 38.8 [22.5, 57.5] & 77.5 [67.2, 86.2] \\
disabled & 1.2 [0.0, 6.7] & 1.2 [0.0, 6.7] \\
\bottomrule
\end{tabular}
\end{table}

\begin{table*}[!t]
\caption{Restored-state control in every weight format, Qwen2.5-7B under
llama.cpp, GSM8K test items 0 to 99: three independent server sessions per format, each serving the same items four times (twice on
the recompute path, twice on the cache-hit path) with the cold state verified
from server telemetry on every request. The sessions serve the same items in
different orders. Counts are given per session; the last two columns count
items whose tokens are identical in all three sessions, out of 100.}
\label{tab:reset}
\centering
\small
\begin{tabular}{@{}lrccccc@{}}
\toprule
& & \multicolumn{2}{c}{path reproduces} & paths & \multicolumn{2}{c}{same in all sessions} \\
\cmidrule(lr){3-4}\cmidrule(lr){6-7}
Weights & items & recompute & cache hit & differ & recompute & cache hit \\
\midrule
F16 & 100 & 100/100/100 & 100/100/100 & 2/2/2 & 100/100 & 100/100 \\
Q8\_0 & 100 & 100/100/100 & 100/100/100 & 43/43/43 & 100/100 & 100/100 \\
Q4\_K\_M & 100 & 100/100/100 & 100/100/100 & 43/43/43 & 100/100 & 100/100 \\
Q3\_K\_M & 100 & 100/100/100 & 100/100/100 & 60/60/60 & 100/100 & 100/100 \\
\bottomrule
\end{tabular}
\end{table*}

\begin{table*}[!t]
\caption{Episode-level divergence in the original grid (production-default
configuration: llama.cpp's server-level prompt cache at its default; vLLM, which
has no such layer, served both cache-enabled passes from one process; on llama.cpp a cache-disabled pass
ran between the two cache-enabled passes, except for Qwen2.5-14B, whose cache-enabled passes ran back to back as
on vLLM). Each configuration ran
the same 80-episode workload twice per arm on one server history, so each row is
a single-history measurement; the replicated sweep of Table~\ref{tab:sweep}
gives the between-history variation for the four Qwen2.5-7B formats under
llama.cpp. Run-to-run divergence, for \emph{cache off} and \emph{cache on}, is the share
of episodes whose trajectory changed when the same arm was repeated; path
divergence compares the two arms. Every cache-off value is
exactly zero. Intervals are the envelope of a circular block bootstrap and the
Wilson interval (Section~\ref{sec:measures}); where serial dependence is weak
the bootstrap is narrower and the Wilson bound is the one shown.}
\label{tab:main}
\centering
\begin{tabular}{lllrrrcrc}
\toprule
& & & & \multicolumn{3}{c}{run-to-run divergence (\%)} & \multicolumn{2}{c}{path divergence (\%)} \\
\cmidrule(lr){5-7}\cmidrule(lr){8-9}
Engine & Model & Weights & $n$ & cache off & cache on & 95\% CI & rate & 95\% CI \\
\midrule
llama.cpp & Qwen2.5-7B & F16 & 80 & 0.0 & 20.0 & [11.2, 30.0] & 36.2 & [26.2, 47.2] \\
llama.cpp & Qwen2.5-7B & Q8\_0 & 80 & 0.0 & 55.0 & [44.1, 65.4] & 61.3 & [50.0, 71.2] \\
llama.cpp & Qwen2.5-7B & Q4\_K\_M & 80 & 0.0 & 77.5 & [67.2, 86.2] & 75.0 & [64.5, 83.8] \\
llama.cpp & Qwen2.5-7B & Q3\_K\_M & 80 & 0.0 & 67.5 & [56.2, 78.8] & 77.5 & [67.2, 86.2] \\
llama.cpp & Llama-3.1-8B & Q8\_0 & 80 & 0.0 & 33.8 & [24.3, 44.6] & 55.0 & [44.1, 65.4] \\
llama.cpp & Llama-3.1-8B & Q4\_K\_M & 80 & 0.0 & 60.0 & [48.8, 71.2] & 70.0 & [59.2, 78.9] \\
llama.cpp & Llama-3.1-8B & Q3\_K\_M & 80 & 0.0 & 62.5 & [50.0, 75.0] & 68.8 & [57.9, 78.8] \\
llama.cpp & Qwen2.5-14B & Q4\_K\_M & 80 & 0.0 & 31.2 & [16.2, 47.5] & 91.2 & [83.0, 96.2] \\
vLLM & Qwen2.5-7B & BF16 & 80 & 0.0 & 8.8 & [2.5, 17.0] & 62.5 & [50.0, 75.0] \\
vLLM & Llama-3.1-8B & BF16 & 80 & 0.0 & 10.0 & [1.2, 20.0] & 53.8 & [41.2, 65.0] \\
\bottomrule
\end{tabular}
\end{table*}

\subsection{RQ3: quantization amplifies divergence}
\label{sec:rq3}

\begin{figure}[!t]
\centering
\includegraphics[width=\columnwidth]{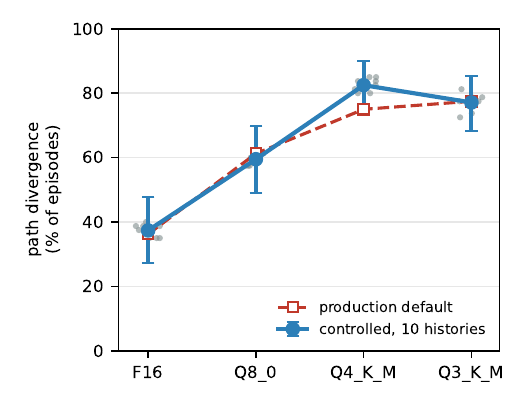}
\caption{Quantized formats widen path divergence: it rises from F16 through
Q8\_0 to Q4\_K\_M, and in the controlled sweep Q3\_K\_M does not exceed Q4\_K\_M. Gray dots are individual random histories; the line gives the
mean over histories with a 95\% interval that resamples episodes and
histories. The dashed line is the production-default grid of
Table~\ref{tab:main}, one history per format. Qwen2.5-7B under llama.cpp, 80
episodes per pass.}
\label{fig:gradient}
\end{figure}

Table~\ref{tab:sweep} and Fig.~\ref{fig:gradient} give the replicated
controlled sweep of Section~\ref{sec:replication}, which is the primary
evidence for this question. Averaged over the ten random histories, path
divergence, the share of episodes whose cache-enabled trajectory departs from
the cache-disabled reference, is \pbv{sweep.F16.mean} percent at F16, \pbv{sweep.Q8.mean} percent at
Q8\_0, \pbv{sweep.Q4.mean} percent at Q4\_K\_M and \pbv{sweep.Q3.mean} percent at
Q3\_K\_M. The between-history standard deviation is only \pbv{sweep.sdmin} to
\pbv{sweep.sdmax} points, far smaller than the differences between formats, so
the aggregate rates were robust to the order in which the episodes were served; most of the
width of the intervals below comes from the sample of episodes. Individual
episodes are less stable: between \pbv{sweep.vary.min} and \pbv{sweep.vary.max}
of the 80 episodes, depending on the format, diverge in some histories and not
in others. Paired by episode and
history, the difference
from F16 is \pbv{sweep.Q8.diff} points at Q8\_0 (95\% CI \pbv{sweep.Q8.difflo}
to \pbv{sweep.Q8.diffhi}), \pbv{sweep.Q4.diff} at Q4\_K\_M (\pbv{sweep.Q4.difflo}
to \pbv{sweep.Q4.diffhi}) and \pbv{sweep.Q3.diff} at Q3\_K\_M
(\pbv{sweep.Q3.difflo} to \pbv{sweep.Q3.diffhi}). Every quantized format lies
above F16 in all 10 histories. The canonical history,
run the same way, gives \pbv{sweep.F16.canon}, \pbv{sweep.Q8.canon},
\pbv{sweep.Q4.canon} and \pbv{sweep.Q3.canon} percent. On the third machine, its F16, Q4\_K\_M and
Q3\_K\_M passes reproduced token for token the controlled
measurements of these cells made earlier with the server-level prompt cache disabled. A second, independently started server
process given the canonical history reproduced the cache-enabled pass in all four formats, in every
one of \pbv{bone.tokens} tokens and their log-probabilities.

The paired trend models agree. Fitted to each random history and averaged over
the histories, the random-intercept model gives an odds ratio of
\pbv{sweep.or.hist} per format step (95\% CI \pbv{sweep.or.cilo} to
\pbv{sweep.or.cihi}, from \pbv{sweep.or.reps} bootstrap replicates that resample
episodes and histories and refit the model). The divergence rate
rises by \pbv{sweep.slope} percentage points per format step (95\% CI
\pbv{sweep.slopelo} to \pbv{sweep.slopehi}, resampling episodes and histories),
and the exact within-episode permutation test gives a two-sided
$p = \pbv{sweep.perm.ptwo}$.

The rise happens in two steps of similar size, from F16 to Q8\_0
(\pbv{sweep.step.Q8minusF16} points, 95\% CI \pbv{sweep.step.Q8minusF16.lo} to
\pbv{sweep.step.Q8minusF16.hi}) and from Q8\_0 to Q4\_K\_M
(\pbv{sweep.step.Q4minusQ8} points, \pbv{sweep.step.Q4minusQ8.lo} to
\pbv{sweep.step.Q4minusQ8.hi}), and both steps are upward in all 10 histories.
Q3\_K\_M adds nothing beyond Q4\_K\_M. It lies below Q4\_K\_M in all 10
histories, by \pbv{sweep.step.Q3minusQ4.abs} points on average, but the interval that also
resamples episodes runs from \pbv{sweep.step.Q3minusQ4.lo} to
\pbv{sweep.step.Q3minusQ4.hi}, so the order of Q4\_K\_M and Q3\_K\_M holds for
this set of episodes but is not established beyond it. We
therefore read the result as an amplification that grows from F16 through
Q8\_0 to Q4\_K\_M, and not as a strict monotone gradient in bit width. The size
of the cache-path perturbation, reported below, follows the same order.
This amplification depends on the implementation. It reflects how the llama.cpp kernels compute with quantized GGUF
weights. Other engines or
quantization schemes may show a different relationship.

\begin{table*}[!t]
\caption{Replicated controlled sweep: cache-enabled against cache-disabled path
divergence for Qwen2.5-7B under llama.cpp, server-level prompt cache disabled
from launch, a fresh server for every pass, 80 episodes per pass. Each random
history is an independent episode order shared by all four formats. The
interval resamples episodes and histories; the difference from F16 is paired
by episode and history.}
\label{tab:sweep}
\centering
\begin{tabular}{lrrrcccc}
\toprule
& canonical & \multicolumn{4}{c}{10 random histories (\%)} & difference & histories \\
\cmidrule(lr){3-6}
Weights & history (\%) & mean & SD & range & 95\% CI & from F16 (points) & above F16 \\
\midrule
F16 & 40.0 & 37.4 & 1.8 & 35.0--40.0 & [27.4, 47.6] & -- & -- \\
Q8\_0 & 60.0 & 59.5 & 1.6 & 57.5--61.3 & [49.1, 69.8] & +22.1 [+9.9, +34.4] & 10/10 \\
Q4\_K\_M & 81.2 & 82.5 & 1.9 & 80.0--85.0 & [74.4, 89.9] & +45.1 [+34.9, +55.6] & 10/10 \\
Q3\_K\_M & 77.5 & 77.1 & 2.5 & 72.5--81.2 & [68.2, 85.4] & +39.8 [+26.9, +52.1] & 10/10 \\
\bottomrule
\end{tabular}
\end{table*}

The production-default grid of Table~\ref{tab:main}, measured under one
history, shows the same rise from F16 to Q4\_K\_M and Q3\_K\_M: path divergence
of 36.2 percent at F16, 61.3 percent at Q8\_0, 75.0 percent at Q4\_K\_M and
77.5 percent at Q3\_K\_M. A random-intercept model on those four cells gives an
odds ratio of \pbv{grid.or} per format step (95\% CI \pbv{grid.orlo} to
\pbv{grid.orhi}; random-intercept standard deviation \pbv{grid.sigma} on the
logit scale). The conditional logistic regression gives \pbv{grid.clogit.or}
(95\% CI \pbv{grid.clogit.lo} to \pbv{grid.clogit.hi}) from the
\pbv{grid.clogit.informative} episodes whose status differs across formats, the
exact within-episode permutation test gives a two-sided $p =
\pbv{grid.perm.ptwo}$, and a bootstrap that resamples episodes jointly across
the four formats puts the increase at \pbv{grid.slope} percentage points per step
(95\% CI \pbv{grid.slopelo} to \pbv{grid.slopehi}). Llama-3.1-8B rises from 55.0 percent at
Q8\_0 to 70.0 percent at Q4\_K\_M, with Q3\_K\_M at 68.8 percent, within a few points
of the cell above it. The 16-bit cell matters most for interpretation: divergence is already present without any
quantization, so quantization is not the cause of the effect. It is a
multiplier.

\begin{figure*}[!t]
\centering
\includegraphics[width=\textwidth]{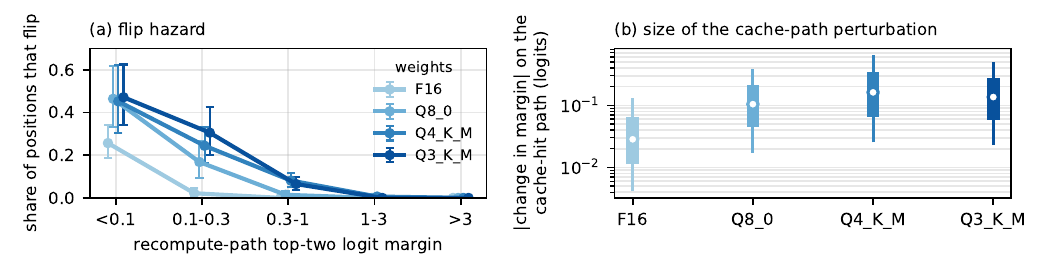}
\caption{Where flips happen and why quantization adds more of them.
(a) Share of positions whose token flips on the cache-hit path, by the top-two
logit margin on the recompute path, with 95\% intervals from a bootstrap over
episodes. (b) Size of the change that the cache-hit path makes to that margin
at positions that did not flip: median, interquartile range and 10th to 90th
percentiles. Canonical history, so every position counts once; Qwen2.5-7B
under llama.cpp.}
\label{fig:margins}
\end{figure*}

Why the quantized formats diverge more is a separate question, and the five
log-probabilities recorded at every position let us test it directly. A
token can flip when the cache path changes the relative logits enough for a
competing candidate to overtake the original winner. Quantization could raise
the flip rate in two ways: by narrowing the top-two gap, so that more positions
sit close to a tie, or by enlarging the change that the cache path makes to the
logits. The two
explanations predict different things, and Fig.~\ref{fig:margins} separates
them. Every divergent episode on the canonical history has its first flip at
a position where both paths had seen the same prompt and the same preceding
tokens, so each flip can be read against the margin the recompute path had
there. The flip hazard falls steeply with that margin in every format
(Fig.~\ref{fig:margins}a, in coarse bins). In finer bins at Q4\_K\_M,
\pbv{mech.Q4.bin0.flips} of the \pbv{mech.Q4.bin0.pos} positions with a margin
below 0.01 logits flipped, \pbv{mech.Q4.bin1.flips} of \pbv{mech.Q4.bin1.pos}
between 0.01 and 0.03, \pbv{mech.Q4.bin2.flips} of \pbv{mech.Q4.bin2.pos}
between 0.03 and 0.1, \pbv{mech.Q4.bin3.flips} of \pbv{mech.Q4.bin3.pos}
between 0.1 and 0.3, \pbv{mech.Q4.bin4.flips} of \pbv{mech.Q4.bin4.pos} between
0.3 and 1, \pbv{mech.Q4.bin5.flips} of \pbv{mech.Q4.bin5.pos} between 1 and 3,
and none of the \pbv{mech.Q4.bin6.pos} positions above 3.

The share of positions close to a tie rises little with
quantization. On the recompute path, \pbv{mech.F16.near}, \pbv{mech.Q8.near},
\pbv{mech.Q4.near} and \pbv{mech.Q3.near} percent of tokens have a margin below
0.1 logits at F16, Q8\_0, Q4\_K\_M and Q3\_K\_M. On per-episode shares, a Page
test for an increase across formats gives $p = \pbv{mech.near.pageinc}$, and the
share at Q3\_K\_M is \pbv{mech.near.ratio} times that at F16 (95\% CI
\pbv{mech.near.ratiolo} to \pbv{mech.near.ratiohi}), far from the change in the
perturbation reported next. What grows is the perturbation. We measure it as the absolute
difference between the two paths' top-two logit margins. We restrict this
measurement to positions in cache-hit requests where both paths saw the same
prompt and preceding tokens and emitted the same token. This measures a change
in the margin rather than in the full logit vector. Its median is \pbv{mech.F16.pert.median} logits at F16, \pbv{mech.Q8.pert.median} at
Q8\_0, \pbv{mech.Q4.pert.median} at Q4\_K\_M and \pbv{mech.Q3.pert.median} at
Q3\_K\_M. Averaged over episodes, the perturbation at Q3\_K\_M is
\pbv{mech.pert.ratio} times that at F16 (95\% CI \pbv{mech.pert.ratiolo} to
\pbv{mech.pert.ratiohi}; Page test for an increase, $p =
\pbv{mech.pert.pageinc}$). It also follows the
divergence rates: it is largest at Q4\_K\_M, the format that diverges most on
this history, and a little smaller at Q3\_K\_M. A larger perturbation also
reaches larger margins. The median recompute-path margin at a flipped position
is \pbv{mech.F16.flipmargin.median} logits at F16,
\pbv{mech.Q8.flipmargin.median} at Q8\_0, \pbv{mech.Q4.flipmargin.median} at
Q4\_K\_M and \pbv{mech.Q3.flipmargin.median} at Q3\_K\_M, and at the same margin
the quantized formats flip more often (Fig.~\ref{fig:margins}a). The extra flips
under the quantized formats are therefore consistent with a larger change in the
top-two margin on the cache path, while the share of decisions close to a tie rises only about
\pbv{mech.near.ratio}-fold.

A final control separates the quantized weight values from the arithmetic that
serves them. We dequantized the Q4\_K\_M weights to a 16-bit file, which keeps
their values to within 16-bit rounding, and served it on the F16 kernels under
four histories, the canonical order and the first three random orders; we ran
this control for Q4\_K\_M only. The result is a model of its own: its recompute path
differs from that of the F16 model on \pbv{bseven.offvsf} of 80 episodes and
from that of the Q4\_K\_M model on \pbv{bseven.offvsq}. Its path divergence,
however, fell to \pbv{bseven.q4deq} percent, against \pbv{bseven.f16} percent for
F16 and \pbv{bseven.q4km} percent for Q4\_K\_M on the same histories. That is
\pbv{bseven.qminusdeq} points below Q4\_K\_M (95\% CI \pbv{bseven.qminusdeq.lo}
to \pbv{bseven.qminusdeq.hi}) and \pbv{bseven.deqminusf} points from F16, with an
interval from \pbv{bseven.deqminusf.lo} to \pbv{bseven.deqminusf.hi} that
includes zero, so no difference in path divergence between the dequantized model and F16
was detected here, although the interval is too wide to establish equivalence. The cache-path perturbation fell with it, to a median
of \pbv{bseven.pert.median} logits against \pbv{mech.F16.pert.median} at F16 and
\pbv{mech.Q4.pert.median} at Q4\_K\_M, while the share of near-tie positions was
\pbv{bseven.near} percent, against \pbv{mech.F16.near} at F16 and
\pbv{mech.Q4.near} at Q4\_K\_M. The amplification therefore comes mainly from the
quantized execution path, the arithmetic llama.cpp uses to serve quantized
weights, rather than from the rounded weight values themselves. In this build that path differs from the 16-bit one in its kernels,
traceable from the dispatch function
\texttt{ggml\_cuda\_mul\_mat} of \texttt{ggml/\allowbreak src/\allowbreak ggml-cuda/\allowbreak ggml-cuda.cu}
at commit 7e4c0a9. In a cache-hit request, the keys and values of earlier
generated tokens were computed one token at a time during decoding. A
recomputation computes the same positions in large batches during prefill. On the RTX~4090,
products with quantized weights use a matrix-vector kernel (MMVQ) for calls of
up to eight tokens. Longer calls use a tiled tensor-core kernel (MMQ). Both
kernels first round the activations to 8-bit blocks. A small difference
upstream can therefore cross a rounding step. With 16-bit weights, the kernel
also changes with the number of tokens. These kernels do not round the
activations to 8 bits. We did not isolate whether the 8-bit rounding or the kernel switch enlarges
the perturbation. A dequantized Q8\_0 control would show whether the first
step of the rise has the same source.

\subsection{RQ4: path divergence appears on both engines}
\label{sec:rq4}

Two independently built engines agree on path divergence. The last
two rows of Table~\ref{tab:main} carry the evidence. Under vLLM, an engine with a
different cache design, different kernels, and chunked prefill enabled, the
cache-disabled arms again repeat exactly, and the cache-enabled arms depart
from them on 62.5 and 53.8 percent of episodes. vLLM returns token strings rather than identifiers on the endpoint we use,
so for this engine ``exactly'' means identical token strings with identical
log-probabilities, at all 30,252 and 33,969 tokens of its two configurations.

The run-to-run rates of the two engines are not comparable. On vLLM at
16-bit (BF16), repeating the cache-enabled arm changed 8.8 percent of episodes,
against 20.0 percent for llama.cpp at F16. The llama.cpp grid, however, ran with the server-level prompt cache at its default and with a
cache-disabled pass between its two cache-enabled passes, the two conditions
that Table~\ref{tab:twobytwo} shows to govern run-to-run divergence on that engine. The vLLM passes ran back to back. We therefore compare the
engines only on whether path divergence appears.

One disclosure belongs with these vLLM figures. The cache-enabled and
cache-disabled arms of the Qwen2.5-7B configuration were launched by different
scripts, and the relaunch set an explicit GPU memory fraction that the original
did not. That parameter governs the block pool the cache draws on, so we
checked whether it mattered: the cache-disabled arm reproduces every token string
and log-probability
across all three sessions, including the differently provisioned ones, which shows that the cache-disabled reference is unaffected by the
provisioning difference. It does not rule out an interaction between that
difference and the cache-enabled path, which is one reason we use the vLLM lane
for the direction of the effect and not its size.

The manipulation check comes from the engine
rather than from the response body. vLLM 0.11.0 does not populate the cached-token field on the
completions endpoint, so every response in the cache-enabled arm reported
no cached-token value even while the cache was working. The engine log resolves
it: prefix caching is recorded as enabled at startup and the reported hit
rate rises from 87.1 to 99.1 percent across 226 logged observations. We note
this because a study that trusted response telemetry alone would have
concluded that the manipulation had failed and discarded a valid arm.

\subsection{RQ5: instability without detectable bias}
\label{sec:rq5}

\begin{table*}[!t]
\caption{Single-turn bridge (GSM8K). Each item is served four times, twice on
the recompute path and twice on the cache-hit path. Both paths are
individually deterministic, yet they disagree with each other on a large
fraction of items once quantized (path divergence, with 95\% Wilson intervals). Correctness
flips occur in both directions. Answers are derived from the stored responses
by numeric comparison; the released artifact reproduces every value.}
\label{tab:bridge}
\centering
\begin{tabular}{llrccrcrrrr}
\toprule
& & & \multicolumn{2}{c}{path determinism} & \multicolumn{2}{c}{path divergence} & flips & \multicolumn{2}{c}{correct} & McNemar \\
\cmidrule(lr){4-5}\cmidrule(lr){6-7}\cmidrule(lr){9-10}
Model & Weights & $n$ & recompute & cache hit & \% & 95\% CI & ($n$) & recompute & cache hit & $p$ \\
\midrule
Qwen2.5-7B & F16 & 200 & 200/200 & 200/200 & 3.5 & [1.7, 7.0] & 0 & 183 & 183 & 1.000 \\
Qwen2.5-7B & Q8\_0 & 200 & 200/200 & 200/200 & 44.5 & [37.8, 51.4] & 4 & 184 & 182 & 0.625 \\
Qwen2.5-7B & Q4\_K\_M & 200 & 200/200 & 200/200 & 41.5 & [34.9, 48.4] & 4 & 180 & 180 & 1.000 \\
Qwen2.5-14B & Q4\_K\_M & 200 & 200/200 & 200/200 & 48.0 & [41.2, 54.9] & 1 & 185 & 186 & 1.000 \\
Qwen2.5-7B & Q4\_K\_M & 500 & 500/500 & 500/500 & 45.4 & [41.1, 49.8] & 12 & 453 & 459 & 0.146 \\
Qwen2.5-7B & Q3\_K\_M & 200 & 200/200 & 200/200 & 48.0 & [41.2, 54.9] & 3 & 178 & 179 & 1.000 \\
\bottomrule
\end{tabular}
\end{table*}

The agentic workload cannot answer the outcome question. Models of this size
solve few of its episodes, from 1.2 percent in the weakest configuration to
18.8 percent for Qwen2.5-14B, so success rates sit near the floor and paired
flips are too few for a precise comparison. Pooled across all ten configurations,
the cache-enabled arm won 12 episodes and the cache-disabled arm won 7. No per-configuration test approaches significance. The largest model we ran
raises the ceiling without changing the conclusion, which suggests the limit
is task difficulty rather than a quirk of the smallest models. We therefore
draw no conclusion about agent task success from these data and report the
trajectory-level results of Sections~\ref{sec:rq1}--\ref{sec:rq4}, which do
not depend on task success, as the agentic contribution.

Outcome conclusions come from the single-turn bridge, where accuracy sits
well clear of the floor at 89 to 93 percent. Table~\ref{tab:bridge} reports
it. Each item
is served four times, twice on the recompute path and twice on the cache-hit
path, so that path divergence is only counted when each path first
agreed with itself. Every path was internally deterministic on every item in
every configuration, which is what licenses the attribution.

The 500-item run at Q4\_K\_M extends the 200-item run at the same setting,
sharing its first 200 problems, so the pooled analysis uses five distinct
configurations: 1300 item-configuration observations over 500 unique problems,
with 20 correctness flips attributable to the cache path. Problems 0 to 199
appear in all five configurations and problems 200 to 499 only in the 500-item
run.

Two patterns appear. The first is a step between 16-bit and every quantized
format. At 16-bit precision the cache path changes the output for 3.5 percent
of items, so at full precision a single-turn request is largely unaffected.
Under quantization the same measurement rises to between 41.5 and 48.0 percent,
and a larger run of 500 items at Q4\_K\_M places the rate at 45.4 percent with
a correspondingly tighter interval. Unlike the agentic sweep, the single-turn
data show no further rise from Q8\_0 to Q4\_K\_M. Comparing the 16-bit rate against the
agentic numbers is instructive: at 16-bit, individual requests diverge on 3.5
percent of items, yet \pbv{sweep.F16.mean} percent of agent episodes do in the
replicated sweep. An episode chains about ten requests, and the gap is
consistent with a single changed token propagating through the remaining
turns.

The second pattern is the more important one for practice.
Correctness flips occur in both directions, and we detected no net change in
aggregate accuracy. Of the 20 flips, 13 favor the cache-hit path and 7 favor recompute. They
fall on 20 different problems, so clustering by problem leaves the inference
unchanged: the problem-clustered sign test gives $p = 0.26$, the same as the
pooled exact McNemar test, and no individual configuration reaches
significance. Six tests were run without
multiplicity correction, so a nominally significant single result would have
warranted caution; none arose. The net change in accuracy on the cache-hit path is +0.46
percentage points, with a 95\% interval of $-0.15$ to $+1.15$ points from a
bootstrap that resamples problems. At these settings the
evidence supports instability rather than degradation.

The measurement here required a correction worth reporting, because it is a
trap for anyone repeating this design. Our collection-time scorer accepted
only the answer format the prompt requested, and the models frequently
answered correctly in a different unambiguous form. Between 13 and 37 percent of
recompute-path responses were scored wrong for formatting alone, which inverted the apparent
accuracy ordering across quantization levels and inflated the flip count
roughly fourfold. Re-deriving every answer offline from the stored response
text, with no new inference, resolves it, and comparing answers numerically
rather than as strings resolves a second, smaller case. Table~\ref{tab:bridge}
reports the corrected figures, and the artifact ships both the original and
the re-derived scores so the correction is auditable. Anyone repeating this
design should validate the extractor against the model's actual output
formats before drawing conclusions from flip counts.

We state the consequence plainly because it is easy to misread in either
direction. Cached serving is not measurably worse at these settings, so a
practitioner choosing caching for throughput gives up no detectable accuracy on
average. What they give up is repeatability: a specific item
answered correctly may be answered incorrectly when an identical request hits
the cache, and the reverse, with no net change detectable in the aggregate. Reporting
only the aggregate hides this, which is the same observation made about
compressed models, where matched average accuracy conceals substantial
per-item disagreement \cite{dutta2024accuracy}.

Concurrent work reaches the same conclusion by a different route. Lorup
\cite{lorup2026stagereplay} compares a retained live cache against a
one-shot prefill of identical tokens on a reasoning benchmark, with a
per-path replica control equivalent in spirit to our four-pass protocol, and
likewise finds substantial suffix divergence, correctness flips in both
directions, and no detectable aggregate accuracy shift. That study adds a bidirectional
cache-transplantation experiment and an FP32 falsification that we do not
have; ours adds a serving-engine setting, a weight-quantization axis, and
multi-turn agent episodes that it does not. We therefore present this
section as independent replication rather than as a new finding, and note
that two studies with different models, benchmarks, precisions, and
execution stacks now agree that the outcome effect is instability without
a detectable direction, in contrast to the systematic bias reported earlier by
Chodavarapu and Xu \cite{chodavarapu2026illusion}.

\section{Discussion}
\label{sec:discussion}

\subsection{What practitioners should take from this}

Enabling the cache changes the guarantee a serving endpoint provides. The
savings are real and large \cite{lumer2026dontbreak}, and nothing in our data
suggests that caching makes models worse on average, so nothing here argues
against prefix caching. The change to the guarantee, however, is documented at most as a brief
caveat \cite{llamacpp2026serverreadme} and is unmeasured in deployment.

Three consequences follow directly. First, a system whose behavior must be
auditable or repeatable, for example one whose outputs are logged for later
review, cannot obtain that property from a cache-enabled endpoint alone,
because the same input is not guaranteed to reproduce the same output.
Second, debugging becomes harder in a specific way: a failure observed once
may not reproduce on replay, not because the input was captured incorrectly
but because the server's cache state differs. Third, an A/B comparison
between two prompts or two models run at different times on a shared
endpoint is confounded by whatever else that endpoint served in between.

The remedies differ by effect. Run-to-run divergence for an identical request
sequence responded to controlling cache state, not to abandoning caching: under
the canonical order, disabling the server-level prompt cache while leaving
prefix reuse on cut it from 38.8 to 1.2 percent, and restarting the server
between runs achieved the same on the second engine. The cost of that control
is small in our setting. On Qwen2.5-7B at Q4\_K\_M, with requests served one at a time and
latencies averaged over three sessions, mean latency per request was
\pbv{lat.default.mean}\,s with the server-level prompt cache at its default and
\pbv{lat.disabled.mean}\,s with it disabled, an increase of
\pbv{lat.overhead.pct} percent. Recomputing every prompt took
\pbv{lat.recompute.mean}\,s. With that cache disabled, prefix reuse kept
\pbv{lat.retained.pct} percent of its latency reduction and stayed
\pbv{lat.speedup} times faster than recomputation. Mean server-reported prefill time per request, which precedes the first
token, rose from \pbv{lat.default.prefill} to \pbv{lat.disabled.prefill}\,ms
because more of each prompt was recomputed; recomputing every prompt took
\pbv{lat.recompute.prefill}\,ms. Within each mode, all three sessions issued the same number of requests, but
the two cache-enabled modes followed slightly different trajectories
(\pbv{lat.req.max} and \pbv{lat.req.min} requests per session), so these are averages over
slightly different request mixes. Restarting the server, which resets all
cache state between runs, took \pbv{lat.default.startup}\,s with the model
files already in the operating-system cache.
These are single-stream measurements; we did not measure throughput under
concurrent load. A workflow that needs repeatable results for an identical
request sequence can pin the cache configuration and reset state at
run boundaries rather than turn prefix reuse off. History dependence under
varying traffic is harder to remove: with the server-level prompt cache
disabled, between
\pbv{sweep.vary.min} and \pbv{sweep.vary.max} of the 80 episodes still changed
divergence status with the history the server had processed, and only isolating each
episode's cache state, as the sentinel control did, removed that dependence;
we did not measure the cost of such isolation. Path divergence, the gap between
the cache-hit and recompute paths, remains under all of these controls. Only
turning prefix reuse off, or using kernels that compute both paths identically
\cite{sglang2025deterministic}, removes it. A more practical middle path, which our per-request telemetry
makes concrete, is to record cache exposure alongside each response so that
irreproducibility is at least detectable after the fact. Only one of the two engines we tested reports this uniformly: llama.cpp
exposes cached tokens on every request, while vLLM 0.11.0 leaves the field
null on the endpoint we used even though the cache is demonstrably active.

The dequantized control has a practical reading as well. At Q4\_K\_M, the extra path
divergence of quantized serving comes mainly from the arithmetic that computes
with quantized weights, so it is a property of the engine and its version as much as
of the model file, which is one more reason to report both. Serving the
dequantized weights on the 16-bit kernels brought path divergence to a level
at which no difference from 16-bit was detected in our measurements, but it gives up the memory saving of
quantization and changes the outputs themselves: the dequantized model's
recompute path differs from the Q4\_K\_M model's on \pbv{bseven.offvsq} of 80
episodes. It is an option for reducing the extra divergence of four-bit
serving rather than a free fix. Path divergence appeared on both llama.cpp and vLLM,
but its amplification by quantization was measured only on llama.cpp with
GGUF weights, and whether the quantized kernels of other engines behave the
same way is not something our data show.

\subsection{Implications for benchmark and agent evaluation}

Published evaluations of language models rarely report cache configuration.
Our results imply that they should. A benchmark score obtained on a
cache-enabled endpoint is a sample from a distribution induced partly by
serving history, not a property of the model and prompt alone, and the
single-turn bridge shows that individual items change correctness at a
measurable rate even when no net change in the aggregate is detected.

The effect is large enough to matter at the scale of differences that
evaluation papers routinely treat as meaningful. Reported variance from
seeds and training noise is generally small \cite{madaan2024quantifying},
and score differences of a few points are commonly discussed as substantive.
Our path-divergence measurements show agent trajectories changing on a majority of
episodes in several configurations, and while trajectory change is not the
same as score change, it is a much larger perturbation than the sources
evaluation practice currently controls for.

This suggests a concrete and cheap addition to existing reproducibility
checklists \cite{pineau2021improving, biderman2024lessons, zhu2025rigorous}:
report the serving engine, its version, the weight format, whether prefix
caching was enabled, and any server-level cache setting such as llama.cpp's
server-level prompt cache, in the same way that decoding parameters are already reported. This
is the kind of configuration-level reporting that critiques of agent
evaluation practice have called for \cite{kapoor2024agents}. Work on
reliability metrics for agents measures repeated-trial variation and
attributes it to the model \cite{yao2024taubench}; our results indicate that
part of that variation may belong to the serving layer instead.

\subsection{Relationship to prior single-turn findings}

The closest prior study established that cached and recomputed decoding
differ numerically under FP16, using single-turn arithmetic, and reported a
systematic accuracy bias across most of its conditions
\cite{chodavarapu2026illusion}. Our single-turn bridge reproduces the
divergence half of that result and, on our models and item set, does not
reproduce the directional half. Both paths are individually deterministic,
they disagree on a substantial fraction of items once quantized, correctness flips occur in
both directions, and we detected no net change in aggregate accuracy.

We read this as a refinement rather than a contradiction. Their setup
differs from ours in models, in item set, and in the absence of a per-path
determinism control, and a directional effect may well be real for their
configuration. The distinction matters for what practitioners should worry
about. A systematic bias would mean cached serving quietly shifts accuracy in one
direction, and would call for correction. Instability without bias means cached serving is
not measurably worse on average but is not repeatable from the request alone, which calls for disclosure and
for care in any workflow that assumes repeatability.

\section{Threats to Validity}
\label{sec:threats}

\textbf{Construct validity.} Our primary measure is divergence in emitted
tokens (identifiers on llama.cpp, strings on vLLM), which is a strict criterion: two runs that produce
semantically identical answers in different words are counted as divergent.
This is deliberate, because the question we ask is whether the system is
reproducible, not whether it is approximately as good. Where the question is
about quality rather than reproducibility we switch to task-outcome scoring
(Section~\ref{sec:measures}) and report task outcomes separately.

\textbf{Internal validity.} The design's main threat is that some
uncontrolled factor, rather than prefix reuse, produces the divergence we
attribute to it. One such factor was present in our first grid and we found
it only by testing for it. The eight llama.cpp configurations of
Table~\ref{tab:main} were collected with the server-level prompt cache at its
default, which inflates the run-to-run measurement through a second layer of
server state; Section~\ref{sec:mechanism} isolates it and reports the corrected
figure. We also verified that the execution order of passes, which differed between
our llama.cpp and vLLM lanes, has no effect for Qwen2.5-7B at Q4\_K\_M once
the server-level prompt cache is disabled; with that cache active, order matters, as Table~\ref{tab:twobytwo}
shows. Both the
original and the controlled measurements are reported rather than one
silently replacing the other. The broader threat is answered by measurement
rather than by argument. Every
configuration in the original grid is executed twice per arm, and the
cache-disabled repetitions are identical throughout, in every emitted token and its log-probability, so no
other source of nondeterminism was observed under our conditions. Batch size is fixed at
one and requests are serial, removing batch-composition effects
\cite{he2025defeating}. Cache precision is held at 16-bit floating point in
all arms so that it never covaries with the weight format. Chunked prefill,
where the engine enables it by default, is left at its default and is
identical across arms of a configuration.

\textbf{A tautology objection.} A reader may respond that reusing cached
state obviously changes arithmetic, so divergence is expected and the
finding is definitional. We accept the mechanism as unsurprising and answer
the objection with an experiment rather than an argument. Restoring the cache
to a known state before each recompute pass makes both paths individually
reproducible, in every weight format and in three independent sessions each,
while the two paths still differ from each other on
\pbv{reset.F16.s1.differ} to \pbv{reset.Q3.s1.differ} of every 100 items,
depending on the format (Table~\ref{tab:reset}). Under the controlled
configurations tested, path divergence is therefore a
function of cache state and not of residual randomness in either path, and cache
state is precisely what deployments do not control. Beyond that we claim the magnitude and the
consequence. Neither the fraction of agent
episodes affected, nor the interaction with quantization, nor the fact that
correctness flips without a detectable shift in aggregate accuracy, follows from
knowing that floating-point addition is not associative. Practice also does
not treat the effect as expected: two production bug reports describe it as
a defect \cite{vllm2025issue18055, vllm2026issue33123}, and published
evaluations rarely report cache configuration.

\textbf{Completeness of the released runs.} Three details are visible in the
artifact and we state them here. The 16-bit bridge log retains 67 records from
an aborted first attempt that was restarted; the analysis keys on item and
pass, so the restarted run supersedes them and the reported values are
unaffected. The earlier controlled measurements of the 16-bit and Q3\_K\_M cells
(Section~\ref{sec:rq3}) were run once rather than twice, so they contribute path-divergence values but no run-to-run
repeat; the replicated sweep of Section~\ref{sec:replication} replaces them as
the evidence for the quantization result. One vLLM cell in the ordering comparison completed without its
end-of-run marker and one was not collected; the reported vLLM figures use the
cells that completed, and the counts are given with each.

\textbf{External validity.} We study open-weight models in the 7 to 14
billion parameter range on a single consumer-class accelerator, because that
is what the study's budget allowed. The two parts of the result
have different reach. Path divergence appears on both engines we tested, while
its amplification by quantization was measured only on llama.cpp with GGUF
weights, since vLLM was run at 16-bit precision. Whether other quantization schemes and
engines show the same amplification is outside what we measured. Behavior at frontier scale, on hosted
endpoints, or under multi-tenant load with cache eviction may differ, and we
make no claim about it. Our workloads are one agentic benchmark and one
mathematics benchmark, both English. The direction we would expect from
these limits is that multi-tenant deployments, where cache contents depend
on other users' traffic, would show more history dependence rather than
less, but we did not measure that and do not assert it.

\textbf{Statistical conclusion validity.} The agentic benchmark is difficult
for models of this size, and their success rates sit near the floor. We
therefore do not draw outcome conclusions from it and say so explicitly in
Section~\ref{sec:results}; the trajectory-level measurements from that
workload remain valid because they do not depend on task success. Outcome
conclusions rest on the mathematics workload, where accuracy sits at 89 to 93
percent, well clear of the floor. Where we report a null result we report the discordant
counts alongside the test, since a p-value on small counts is easy to
over-read \cite{miller2024error}.

\section{Conclusion}
\label{sec:conclusion}

Prefix caching is enabled by default in the serving stacks that most
deployments use, and it is treated as a transparent optimization. This paper
shows that it is not transparent. Holding decoding, seed, batch size, and
request order fixed, we found that repeated execution of an identical
workload reproduced every emitted token and its log-probability when prefix
reuse was disabled, in all ten configurations across two independent engines,
and did not when prefix reuse was enabled in the production-default
configuration. Three further experiments locate the
dependence. Once llama.cpp's
server-level prompt cache is disabled, the order in which passes run has no
effect in the configuration tested. This dependence reflects cache state
carried between runs. Under the canonical episode order, changing the
prompt-cache flag alone moves run-to-run divergence by 37.5 percentage points.
With that cache at its default, run-to-run divergence ranges from
\pbv{bsix.min} to \pbv{bsix.max} percent under five random orders.
When cache state is restored to a known point, both the cached and the
recompute paths reproduce exactly, while continuing to differ from each
other, in every weight format and in independent server sessions. In the
configurations tested, restoring cache state eliminated all observed variation
within each path; the problem is that cache state is invisible to the request,
reported at most as a token count in the response, and uncontrolled by default.

Path divergence appeared on both llama.cpp and vLLM. On llama.cpp, its
magnitude depends on how the model is quantized, and it is larger for
chained requests than for single ones. For Qwen2.5-7B on llama.cpp with GGUF
weights, serving quantized weights widens
path divergence, from \pbv{sweep.F16.mean} percent of agent episodes at 16-bit
to \pbv{sweep.Q4.mean} percent at four-bit, averaged over ten independent server
histories; the rise runs from F16 through Q8\_0 to Q4\_K\_M, and Q3\_K\_M does
not go further. The recorded logit margins point to why: with quantization the
share of decisions near a tie grows only about \pbv{mech.near.ratio}-fold, while
the change that the cache path makes to the top-two margin grows
\pbv{mech.pert.ratio}-fold. Serving the four-bit weights
through the 16-bit kernels left no detectable difference from the 16-bit level
in path divergence, and the dequantized model's median margin change was close to the 16-bit value
(\pbv{bseven.pert.median} against \pbv{mech.F16.pert.median} logits), so at
four-bit the amplification comes mainly from the quantized execution path
rather than from the rounded weights. Single-turn requests
that are nearly unaffected at full precision diverge on more than forty percent
of items once quantized. Chained requests enlarge the effect again, since a
single changed token can propagate through every subsequent turn. Whether other
quantization schemes and kernels on other engines show the same amplification
is the next question to test.

What the effect does to task outcomes is narrower than one might assume.
Individual answers change correctness in both directions, and we detected no
net change in aggregate accuracy. Cached serving is not measurably worse. It is not repeatable from the request alone.
Those are different problems requiring different responses: accuracy
degradation would call for a correction; irreproducibility calls for
disclosure.

The immediate recommendation is cheap. Evaluations and deployment records
should state the serving engine, its version, the weight format, whether
prefix caching was enabled and, on llama.cpp, the server-level prompt-cache
setting, alongside the decoding parameters that are already reported as a
matter of course. Engines should report cache exposure per request
uniformly, which one of the two we tested does and the other does not, so
that irreproducibility is at least detectable after the fact. Longer term,
deterministic modes that cover cached execution already exist in at least
one stack \cite{sglang2025deterministic}, and the underlying mechanism is
understood \cite{he2025defeating, yuan2025numerical}. What our measurements
add is the size of the gap those modes are closing, in the default
configuration that most deployments and most published evaluations
actually run \cite{pape2026silent}, where agent results are often credited
to the model alone \cite{harnessbench2026} and temperature-zero inference
is often treated as reproducible \cite{chu2026margingate}.

\section*{Data Availability}

The measurement harness, the complete raw per-request logs, and the analysis
pipeline are available at \url{https://github.com/aditi-p31/cache-divergence-study}
(tag \texttt{resubmission-2026-10}). The release cited here is archived at
\url{https://doi.org/10.5281/zenodo.23633821}. The logs record, for every request,
the prompt hash, the emitted tokens (identifiers on llama.cpp, strings on
vLLM) with their log-probabilities, the latency and, on llama.cpp, the
cached-token count reported by the server. Every number and figure in this paper is regenerated from those logs and the
released telemetry summary by the scripts in \texttt{analysis/} and \texttt{harness/}: \texttt{reextract\_\allowbreak bridge.py} derives the
single-turn answers; \texttt{analyze.py} and \texttt{analyze\_\allowbreak repair.py} write
\texttt{findings.json} and \texttt{repair\_\allowbreak findings.json};
\texttt{robustness.py} computes the position-dependence test of
Section~\ref{sec:mechanism}; \texttt{results/cache\_\allowbreak telemetry.json} holds the vLLM cache hit
rates that \texttt{harness/extract\_\allowbreak cache\_\allowbreak telemetry.py} read from the engine log;
\texttt{revision\_\allowbreak stats.py} computes the dependence-aware intervals, the paired
trend models and the problem-level GSM8K analysis; \texttt{phaseb\_\allowbreak stats.py}
analyzes the replicated sweep and its controls, including the top-two margin
perturbation and the latency measurements; \texttt{or\_\allowbreak bootstrap.py} computes
the bootstrap interval of the replicated-sweep odds ratio in Section~\ref{sec:rq3}; \texttt{make\_\allowbreak phaseb\_\allowbreak values.py}
writes the values from the replicated sweep and its controls that the text quotes; and \texttt{make\_tables.py}, \texttt{make\_figures.py},
\texttt{make\_\allowbreak phaseb\_\allowbreak tables.py} and \texttt{make\_\allowbreak phaseb\_\allowbreak figures.py} produce the
tables and figures. The episode orders, their seeds and the run orchestration
for the replicated sweep are included. Pinned engine versions, model checksums, and the
launch flags used for each configuration are recorded in the repository. They
are needed to reproduce the results, because the measured effects depend on the
engine version and the cache configuration.

\section*{Acknowledgment}

The author used Claude (Anthropic) to assist with measurement and analysis
code, figure generation, and manuscript preparation. The author designed the
study, verified all reported results against the raw data and primary sources,
and takes full responsibility for the content. All hypotheses, interpretations,
and conclusions presented in this study reflect the author's original ideas.

\bibliographystyle{IEEEtran}
\bibliography{refs}

\begin{IEEEbiography}[{\includegraphics[width=1in,height=1.25in,clip,keepaspectratio]{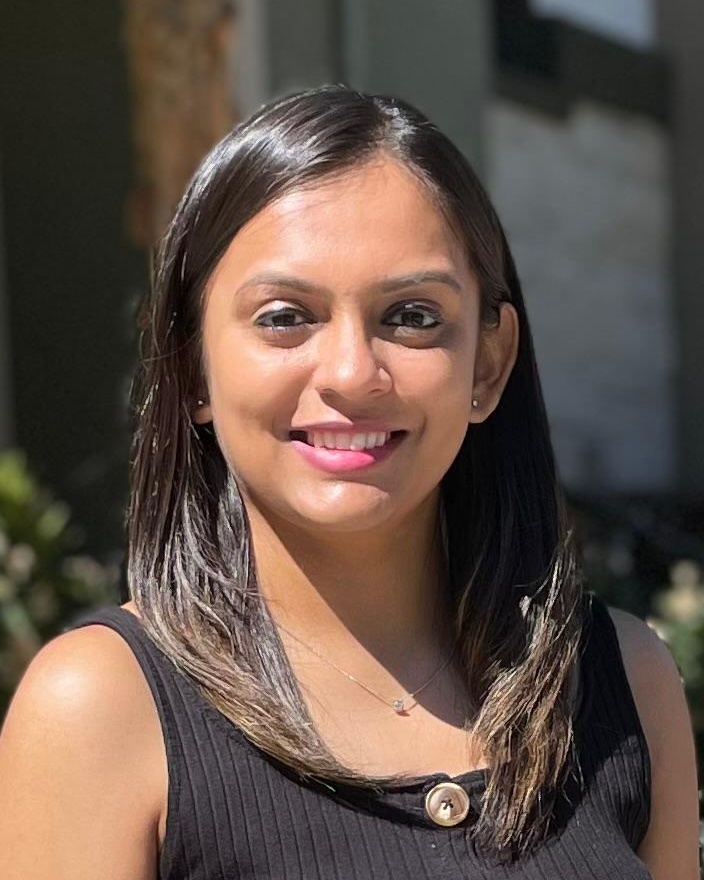}}]{ADITI PATODIYA}
(Senior Member, IEEE) received the B.Tech. degree in computer engineering
from Charotar University of Science and Technology, India, and the M.S. degree in computer science from California State University, Long Beach, CA, USA. She
is a Senior Software Engineer based in California with over 10 years of
industry experience, a reviewer for computing venues including CHI,
Supercomputing, ECIS, and ISMAR, and a technical writer. Her expertise spans
enterprise AI infrastructure, context engineering, and software engineering
for large language model systems. Her work focuses on designing and
architecting scalable backend infrastructure and context pipelines capable of
handling large-scale production traffic. Her current research centers on LLM
inference reproducibility, prompt caching behaviors, and the measurement of
agentic systems in production deployments.
\end{IEEEbiography}

\global\eoddefinedtrue

\end{document}